\documentclass[compsoc, conference, letterpaper, 10pt, times]{IEEEtran}
\usepackage{url}
\usepackage[usenames,dvipsnames]{color}
\usepackage{graphicx}
\usepackage{subfigure}
\usepackage{balance}
\usepackage{multirow}
\usepackage{lipsum}
\usepackage{xspace}
\usepackage{booktabs}
\usepackage{times}
\usepackage{mdwlist}
\usepackage{etoolbox}
\usepackage[hyphenbreaks]{breakurl}
\AtBeginEnvironment{thebibliography}{\linespread{0.98}\selectfont}

\begin{document}

\title{{SoK: An Analysis of Protocol Design:\\
      Avoiding Traps for Implementation and Deployment}
}

\author{
    \IEEEauthorblockN{Tobias Fiebig\IEEEauthorrefmark{1}, Franziska
Lichtblau\IEEEauthorrefmark{1}, Florian Streibelt\IEEEauthorrefmark{1}, \\Thorben
Kr\"uger\IEEEauthorrefmark{1}, Pieter Lexis\IEEEauthorrefmark{2}, \\Randy
Bush\IEEEauthorrefmark{3}, Anja Feldmann\IEEEauthorrefmark{1}}
    \IEEEauthorblockA{\IEEEauthorrefmark{1}TU Berlin
    \\\{tobias, franziska, florian, thorben, anja\}@inet.tu-berlin.de}
    \IEEEauthorblockA{\IEEEauthorrefmark{2}PowerDNS.COM BV
    \\pieter.lexis@powerdns.com}
    \IEEEauthorblockA{\IEEEauthorrefmark{3}Internet Initiative Japan
    \\randy@psg.com}
}


\newcommand{\at}[0]{\textasciitilde}
\newcommand{\kv}[0]{key-value store\xspace}
\newcommand{\kvs}[0]{key-value stores\xspace}
\newcommand{\Kvs}[0]{Key-value stores\xspace}
\newcommand{\net}[0]{Internet\xspace}
\newcommand{\mks}[0]{misconfigurations\xspace}
\newcommand{\Mks}[0]{Misconfigurations\xspace}
\newcommand{\mk}[0]{misconfiguration\xspace}
\newcommand{\mko}[0]{misconfiguration opportunities\xspace}
\newcommand{\Mk}[0]{Misconfiguration\xspace}
\newcommand{\Mko}[0]{Misconfiguration Opportunities\xspace}
\newcommand{\mked}[0]{misconfigured\xspace}
\newcommand{\Mked}[0]{Misconfigured\xspace}

\newcommand{\early}[0]{\textbf{\emph{Early Internet}}\xspace}
\newcommand{\emerging}[0]{\textbf{\emph{Emerging Threats}}\xspace}
\newcommand{\complex}[0]{\textbf{\emph{Complex Security}}\xspace}
\newcommand{\simplicity}[0]{\textbf{\emph{A new Simplicity}}\xspace}

\newcommand{\Early}[0]{\textbf{\emph{Early Internet}}\xspace}
\newcommand{\Emerging}[0]{\textbf{\emph{Emerging Threats}}\xspace}
\newcommand{\Complex}[0]{\textbf{\emph{Complex Security}}\xspace}
\newcommand{\Simplicity}[0]{\textbf{\emph{A new Simplicity}}\xspace}

\newcommand{\theoretical}[0]{``theoretically broken''\xspace}
\newcommand{\feasible}[0]{``feasible to break''\xspace}

\newcommand{\good}[0]{good enough\xspace}
\newcommand{\Good}[0]{Good enough\xspace}
\newcommand{\perfect}[0]{perfect security\xspace}
\newcommand{\Perfect}[0]{Perfect security\xspace}

\newcommand{\details}[0]{security features\xspace}
\newcommand{\Details}[0]{Security Features\xspace}
\newcommand{\cia}[0]{security properties\xspace}
\newcommand{\Cia}[0]{Security Properties\xspace}
\newcommand{\docs}[0]{support\xspace}
\newcommand{\Docs}[0]{Support\xspace}
\newcommand{\pub}[0]{publications\xspace}
\newcommand{\Pub}[0]{Publications\xspace}
\newcommand{\usage}[0]{visible instances\xspace}
\newcommand{\Usage}[0]{Visible Instances\xspace}

\newcommand{\yes}[0]{$\bullet$\xspace}
\newcommand{\no}[0]{$\circ$\xspace}
\newcommand{\na}[0]{-\xspace}

\newcommand{\comment}[1]{}
\newcommand{\todo}[1]{\textcolor{blue}{ #1}}
\newcommand{\fixme}[1]{\textcolor{red}{\textit{#1}}}
\newcommand{\tfiebig}[1]{\textcolor{PineGreen}{TF: \textit{#1}}}
\newcommand{\fl}[1]{\textcolor{Orange}{FL: \textit{#1}}}
\newcommand{\fls}[1]{\textcolor{Blue}{FS: \textit{#1}}}

\newcommand{\hdsec}[3]{
\noindent\rule{\columnwidth}{0.4pt}

\noindent
\textbf{Section Description:}

\noindent
Length: \textbf{#1 p}, \\
Author(s): \textcolor{red}{\textbf{#2}} \\
Content-Description: 

\begin{itemize}
#3
\end{itemize}

\noindent\rule{\columnwidth}{0.4pt}
}

\newcommand{\hdsubsec}[3]{
\noindent\rule{\columnwidth}{0.4pt}

\noindent
\textbf{Subsection Description:}

\noindent
Length: \textbf{#1 p}, \\
Author(s): \textcolor{red}{\textbf{#2}} \\
Content-Description: 

\begin{itemize}
#3
\end{itemize}

\noindent\rule{\columnwidth}{0.4pt}
}

\maketitle

\begin{abstract}
Today's Internet utilizes a multitude of different protocols. While some of 
these protocols were first implemented and used and later documented, other 
were first specified and then implemented. 
Regardless of how protocols came to be, their definitions can contain traps 
that lead to insecure implementations or deployments. A classical example is 
insufficiently strict authentication requirements in a protocol specification.
The resulting \Mks, i.e., not enabling strong authentication, are common root causes for
Internet security incidents.  
Indeed, Internet protocols have been commonly designed
without security in mind which leads to a multitude of \mk traps. While this
is slowly changing, to strict security considerations can have a similarly bad
effect. 
Due to complex implementations and insufficient documentation, security features
may remain unused, leaving deployments vulnerable.

In this paper we provide a systematization of the security traps found in common 
Internet protocols. By separating protocols in four classes we identify major factors
that lead to common security traps.
These insights together with observations about end-user centric usability and
security by default are then used to derive recommendations for improving existing
and designing new protocols---without such security sensitive traps for operators,
implementors and users.

\end{abstract}

\section{Introduction}


Security incidents involving Internet services have become regular
events.  Examples include: (a) disclosures of information, e.g.,
petabytes of personal data stored in unprotected key-value stores -
NoSQL databases ~\cite{mongodbsec,binaryedge2015a}.  (b) Unauthorized
access via the Internet to systems, e.g., supervisory control and data
acquisition systems (SCADA)~\cite{meixell2013out} which are used to
control systems which range from light installations to oil platforms.
(c) Undesired publication of information, e.g., health data from the
United Kingdom on an HTTP root~\cite{wired2014a}, or an UEFI signing key
on an internal FTP server which was available via the
Internet~\cite{caudill2013a}.

While programming vulnerabilities are well-studied,
e.g., see~\cite{mcgraw2006software,mcgraw2004software}, insufficient attention
has been paid to poorly designed and hard to configure protocols.  We
focus on protocol security practices with their varying na\"ivete,
complexity, and weaknesses.  In the context of this paper, we use the term
protocol to refer to transport and application layer communication
protocols.  We use the term service to refer to a deployed instance that
is offering the service associated with the protocol.

Indeed, among the root causes of the above severe incidents are \mks.
While this is, in principle, well known, e.g.,~\cite{cuppens2005detection,xu2013not},
to date, the main and only explanation has been human error. We claim
that there are more fundamental reasons for such \mks which stem from
the design of the Internet's protocols themselves, the security
assumptions or configuration choices offered by protocols. These lead to
services which are prone to \mk.

Our study complements the multitude of individual incidents documented
in the scientific literature and the anecdotes in systems lore with a
macroscopic systematic survey of Internet protocols and their
corresponding services.  More precisely, we investigate which of the
underlying assumptions during protocol design lead to \mks during
service deployment.  We refer to this as \mk prone protocols/services.

In the context of this paper we consider the following security relevant
\mks. A service is subject to \mk if it is deployed on the Internet in such a
way that any of the three main security properties - confidentiality,
integrity, or availability (CIA) - can be tainted.

There are many reasons for \mks: (a) the operator does not follow best
practices regarding network settings by which the service is deployed, (b) the
operator does not use the default configuration settings leading to tainted
CIA, or (c) the operator uses the default configuration settings and they lead
to tainted CIA.

Given the newest Internet trends, services in the cloud, Internet of
Things including Industry~4.0, autonomous systems, e.g., self-driving
cars, and mobile applications, we expect even more diversity and
complexity and, thus, more security incidents. Considering the 
link between mismanagement and
maliciousness~\cite{zhang2014mismanagement,liu2015cloudy}, many of these
incidents will involve \mks. Thus, we claim that a systematic review of
why \mk occurs is needed.

\textbf{Contributions:} We review the assumptions under which protocols have
been designed along two dimensions, (a) the assumed strength of the attacker -
weak vs.\ strong - and (b) the defense paradigm - good enough vs.\ perfect
security. By using these dimensions to group protocols we find four major
clusters, one in each quadrant.  We name these clusters \early for weak
attacker/good enough, \emerging for weak attacker/perfect security, \complex
for strong attacker/perfect security, and \simplicity for strong attacker/good
enough. The names capture the design essentials as well as the mindset of the
protocols in each of the classes. Furthermore, the names express how the
security mindset of protocol design evolved over time.  From this
systematization of \mk prone protocols we derive a set of specific action
items. To keep protocols secure and \mk resilient these must be consider when
introducing new or updating old protocol specifications.

\textbf{Structure:} This paper is structured as follows: We introduce our
systematization methods in Section~\ref{sec:method}. Next, we dedicate a
chapter to each of the four classes, Sections~\ref{sec:c1} to~\ref{sec:c4},
examining how and why protocols and services in the specific class are prone to
\mk. For each class we show its properties based on a few examples which we
discuss in detail. Each of these chapters concludes with a discussion which
summarizes our observations about the \mks within the class.  Finally, in
Section~\ref{sec:conclusion}, we provide design recommendations for future
protocols as well as deployed services, and end with concluding remarks.

\section{Systematization Method}
\label{sec:method}

Thousands of Internet protocols have been proposed, developed, and
deployed on the Internet; therefore an exhaustive analysis is beyond
the scope of this paper. Rather we extract essential features from a
subset of \mk prone protocols and use
these to derive our systematization.

\subsection{Example protocol selection}
Our choice of protocols is driven by the following considerations. First, we
choose protocols that are commonly used and/or are new and upcoming. Next, we
choose some were \mks can potentially have a large impact or those where
system lore states that they are easily \mked.  We augment this list by
protocols that capture corner cases. 
From these we selected a set of protocols that are most iconic vor the relevant
class. Due to the size limitations of the work at hand, only a subset can be 
introduced in detail and displayed in Table~\ref{tbl}.
We focus on the server, not client side. Thus,
pure end-user focused protocols as well as client \mks are beyond the
scope of this paper.

\subsection{Security relevant \mks}
If a service is deployed in such a way that its CIA is tainted because
of a \mk we call the \mk security relevant. There are three main reasons
for such \mks. The first is when the operator of the service does not
follow best common practice (BCP). The second is when the operator uses their
own configuration which taints the CIA of the service. The third is
when the operator uses the default configuration, but the CIA is still
tainted, likely due to incorrect defaults.

We refer to \mk prone protocols and services. But in the end it is the
service that is \mked, often facilitated by design choices in the
protocol. Examples for common \mks are that the service is deployed in a
network setting which deviates from the one for which it, or its
default configuration, was designed.

\subsection{Security guidelines for protocol design}

Request For Comments (RFCs) document Internet protocols and services. Since
1992, each RFC must address the topic of security, according to, IETF processes,
(RFC1311~\cite{rfc1311}, RFC1543~\cite{rfc1543}, RFC2223~\cite{rfc2223},
RFC7322~\cite{rfc7322}), which document what an RFC must contain. Indeed,
RFC1311~\cite{rfc1311} states: \emph{``All STD RFCs must contain a section that
discusses the security considerations of the procedures that are the main topic
of the RFC.''}

Over time, the community has realized that this statement by itself is
not sufficient and the specification of the security requirements have
gotten stricter, see RFC3552~\cite{rfc3552} from 2003, the Best Current
Practices for \emph{``Writing RFC Text on Security Considerations''}.
The goal of this requirement is to make all protocol designers and
implementers aware of possible security implications. Given that this
basic requirement existed in 1992, we conclude that the importance of
security has been recognized for at least 23 years.

\subsection{Review of security threats for protocols and services}

One of the motivations for including a security section in each RFC is to make
the protocol designer consider the following two questions: (a) against whom to
defend and (b) how to defend.


We find that protocol designers consider different kinds of attackers, ranging
from very weak to very strong.  The weak attacker is either
unskilled or is resource limited. The strong attacker is very skilled and has
all necessary resources in their hands.

Defining how to defend is more difficult as it depends on the eyes of
the beholder. Some argue that the cost of breaking security should be
larger than the value of the protected asset.  This goes back to
Pfleeger and Pfleeger~\cite{pfleeger2002security} and specifies that
one should put up a wall against threats at least high enough that most
attacks will not break the wall. Moreover, the wall should not cost more
than the protected asset. We call this the ``\good'' approach to
security. Others, especially the field of
cryptography~\cite{kramer2015cryptography}, follow the approach of
``\perfect''.  \Perfect refers to using every possible mean to achieve
security. We refer to these two approaches as the defense paradigm.

\subsection{Classification of protocols and services}

Thus, we have two-dimensions, namely, the capabilities of
the attacker and the defense paradigm. Using these dimensions
we classify our selected set of protocols and identify four major
clusters. These clusters correspond to the four quadrants of the two dimensional
space. We refer to them as: \early, \emerging, \complex, \simplicity.

\begin{description}
\item[Weak attacker - \good] ~\newline This class contains those
  protocols that are designed for a friendly, collaborative environment - the
  Internet when security was not yet a major concern. This class, in
  particular, contains those protocols that initially were designed 
  under the assumptions that there is no attacker, or still carry artifacts from
  that idyllic time. Such an attacker is the weakest one possible.
  We refer to this class as \early.
\item[Weak attacker - \perfect] ~\newline This class no longer assumes that
  the environment is entirely friendly. Rather it recognizes that there are
  threats, but not yet by sophisticated attackers. However, since significant
  assets can be at stake, even attacks that are only theoretically possible are
  considered.  We refer to this class as \Emerging.
\item[Strong attacker - \perfect] ~\newline This class captures the
  protocols that consider security a necessity at all costs. As a result, the
  protocols in this class are designed to handle strong attackers and be safe
  against all, even theoretically conceived, attack vectors.
  However, as a result, they are often complex and hard to deploy, maintain, and
  difficult to use. We refer to this class as \Complex.
\item[Strong attacker - \good] ~\newline This class contains those
  protocols who's designers recognize that strong attackers exist but
  also value protocols that ``just work'' out of the box. Therefore, the
  designer does not try to defend the asset against every possible
  attack by reducing the attack surface.  In this class we see a
  conscious choice between security and operational ease, favoring the
  latter.  We refer to this class as \Simplicity.
\end{description}

Interestingly, when one considers when most protocols in each of the
above classes were designed, we find that Internet protocol designers
have started with protocols in the \early category, and moved to ones
from \emerging when they realized that the Internet was no longer
nice. However, while the core ideas did not change, the protocols were
hidden behind fences such as DMZs. Since this did not suffice, they
moved to \complex.  As these were hard to maintain or difficult to
use, we see a new trend towards \simplicity.

\subsection{Systematization}
\setcounter{paragraph}{0}

In Sections~\ref{sec:c1}--\ref{sec:c4} we take a closer look at each of the
above classes. For each class, we identify a set of representative protocols
which we analyze according to five sets of features.

\noindent{\bf{\Details:}}
Among the essential security features that protocols should support are
authentication, authorization, and use of encryption (TLS for transport layer
encryption).  We mark for each
protocol/service which of these features is (a) in common use (\yes), (b)
implemented but not commonly used (\no), (c) not implemented  (\na). If the
protocol does not support a security feature we leave the space blank.

\noindent{\bf{\Mk traps:}} We consider the possible \mk traps which a
protocol/service may have.  Under \textbf{NoAuth} we capture if (a)
authentication is not offered by the protocol, (b) typically not
implemented, or (c) typically not configured in the deployed
service. Under \textbf{Credentials} we note if the service is commonly
deployed with weak default credentials. Under \textbf{Artifacts} we note
if protocol features common at design time lead to \mks when used
today. With \textbf{Fencing} we note those cases that depend on
firewalling, etc. to ensure that they are not reachable from the
Internet.  We mark those protocols \textbf{NoUse} that are hardly deployed even
though they replace earlier protocols, e.g., prior versions, with major \mk traps. 

\noindent{\bf{\Docs:}}
As \mks often occur due to poor technical support for operators, we
look closely at that for each of our representative protocols.
More precisely, we look at the documentation for securely deploying the
service and check if it is mostly (\no) or always (\yes) 
misleading, lacking, too complex, or
otherwise insufficient.  Another common aspect that leads to \mks
are bad defaults.  But rather than looking at what can go wrong we check if the
defaults are always (\yes) or sometimes not ``sane'' (\no).  Sane in the sense that 
they enable or enforce enabling the supported security features of the 
protocol/service to ensure confidentially, integrity, and availability by default.

\noindent{\bf{\Pub:}}
Here we capture if problems that can lead to \mks discussed under \mk
traps or \docs are already well known either in the academic world or
the security community.  If it is known in the academic world, we refer
to a representative paper. If it is known in the community we note
approximately when it became part of the system engineering lore.

\noindent{\bf{\Usage:}} Next we try to estimate the number of systems,
or rather IP addresses, that offer the service/protocol under
discussion. We rely on different data sources, among them (a) publicly
available data sets or representative papers, (b)
zMap~\cite{durumeric2013zmap} scans by the authors, and (c) search results
from Shodan~\cite{bodenheim2014evaluation,shodan2015a}. Shodan is ``the
world's first search engine for Internet-connected devices''. While
Shodan does provide an estimate of the possible number of IPs offering a
service it is neither complete, covers all available
ports~\cite{bodenheim2014evaluation}, nor are all instances per se
vulnerable.

\section{The Early Internet}
\label{sec:c1}
\setcounter{paragraph}{0}

This class includes those protocols that were designed in the context of
the early Internet, roughly from 1960 -- 1988, where attacks had yet to
be considered and therefore protocols were designed without security
considerations.  The paper by David Clark about ``The Design Philosophy
of the DARPA Internet Protocols''~\cite{clark1988design} does not even
contain the term security even though availability in the sense of
survivability is a major goal. After all, the main goal of the Internet
was to interconnect existing networks with the implicit assumption that
all participants worked towards the common goal of communication.

As a result attacks were not yet common. So, the need for security
either did not exist or was extremely limited. Towards the end of the
era, attacks against operating systems became more prominent and the
first major Internet worm, namely the Morris worm,
was let loose~\cite{orman2003morris,spafford1989internet}.

Most popular protocols from that era have been updated to remain usable in
today's hostile Internet. However, this does not remove all \mk
traps. Today many still have artifacts of  their design for a friendly
Internet. 

Thus, the threat model of this class is: ``weak attacker'' with ``\good''.  The
representative protocols we examine are: SMTP and DNS.
Other protocols in this class include: TFTP, FTP, Finger, rexec, Chargen, NIS, RIP, NTP,
WHOIS, Ident, XDMCP/X11, Syslog, rsync and IRC.

\subsection{Example Protocols}
\setcounter{paragraph}{0}

\smallskip
\noindent{\bf{FTP:}}
The file transfer protocol (FTP) is an application layer protocol for
Internet file
transfer between hosts. \comment{Its \emph{``primary function is to
facilitate transfer of files between hosts, and to allow convenient use of
storage and file handling capabilities of other hosts.''}
}
FTP is one of the earliest Internet protocols and was first documented
by RFC 114~\cite{rfc114} in 1971. It provides authentication within the protocol and
authorization via the operating system's file system access controls.
It does not offer encryption. However, it can be used over a TLS tunnel,
see RFC4217~\cite{rfc4217} from 2005.

The major \mk pitfalls for FTP servers are related to either missing
authentication or insufficient authorization and directory limits.
These are:
(a) Enable anonymous logins to share files publicly, see RFC1635~\cite{rfc1635}. 
Hence, files uploaded to a server that allows anonymous access
become public. Recent examples show that, e.g., private  keys~\cite{caudill2013a} 
can be exposed this way.
(b) If, in addition,  write access is enabled files can be deleted and/or 
overwritten.  The FTP servers can also be abused to share malicious content. 
This issue has been  discussed as common example by Uppuluri and 
Skar~\cite{uppuluri2001experiences}.
(c) Faulty configuration of the root-directory of an FTP server may
expose system files. If, e.g., a UNIX machine
exposes its global root directory, FTP users can access all files that are
accessible to the user operating the FTP server. 
(d) FTP's dedicated data-channel enables an attacker to
%
%
send files containing service commands to a remote server, e.g., SMTP.  
This attack can be used for amplification and firewall evasion~\cite{helmer2002software}.  
RFC2577~\cite{rfc2577} recommends disallowing data-channel connections to low-ports as 
mitigation.  



To counteract some of the above threats, some modern FTP implementations, such
as vsFTPd, ship a systematically locked down default configuration. It requires
extensive user action to enable anonymous, write, and, non-directory-restricted
access. The documentation of vsFTPd is short and precise.
Other widespread FTP implementations, e.g., the versatile solution ProFTPd 
have a more complex documentation due to their larger feature set. 

\smallskip
\noindent{\bf{TFTP:}}
\label{sec:tftp}
The Trivial File Transfer Protocol (TFTP) is a ``very simple protocol used to
transfer files'' and is documented in RFC783~\cite{rfc783} dated 1981.  TFTP only supports
reading and writing files and lacks most of the advanced features of
FTP.  It uses UDP as transport layer protocol.  TFTP is often used for bootstrapping
by providing access to files needed for system boot such as boot images, firmware
updates, or network device configurations files.  Revision-2,
RFC1350~\cite{rfc1350}, fixed the ``Sorcerer's Apprentice'' protocol
bug - a major data retransmission problem which leads to packet amplification.

TFTP itself does not provide authentication or authorization.  It does offer
limited protection by means of the operating system's file system access
controls. Modifying files on an TFTP server can be restricted by most
implementations.  TFTP does not support encryption.
TFTP is by design insecure which has also led to its most common use case, to
allow bootstrapping of unprovisioned systems that therefore have no
security credentials.

TFTP is also among the first protocols to suffer from unintended
amplification attacks, see RFC1350~\cite{rfc1350}.  This is one of
the earliest amplification attacks of stateless protocols. 
TFTP servers, unless shielded from general access, are subject to disclosure
attacks~\cite{rfc4732}. Indeed, they often enable transitive attacks, in which
an attacker first retrieves the confidential configuration files, including 
encryption, authentication, and authorization secrets. Then the attacker uses
this information to access the systems newly configured over TFTP.

If the TFTP server allows write access, attackers can, in principle, overwrite any
of the configuration files with their own, resetting passwords or configuring additional known credentials for super users. Similarly, the attacker
can alter the available system images - boot as well as full system images - to
include backdoors. Once provisioned they enable the attacker to take over the
corresponding systems.


The documentation of the most common TFTP server implementations is short and
precise and includes a sensible discussion of the security issues. Moreover,
the suggested configurations are appropriate.

To counteract most of the above threats, almost all TFTP servers must be
isolated from public access and subject to strict access rules or
firewalls. Thus, the main \mk trap is missing fences. Indeed, in 2014,
MacFarlane et al.~\cite{macfarlane2015evaluation} found more than 600,000
publicly accessible TFTP servers. However, fencing is insufficient against
inside attackers.

\smallskip
\noindent{\bf{DNS:}}
\label{sec:dns}
Since the early 1980's the Domain Name System (DNS), RFC882 and
RFC883~\cite{rfc882,rfc883}, is used to map hostnames to IP addresses and vice versa.

DNS is a hierarchical distributed database organized in 
independently administered DNS zones. These zones are implemented as subtrees in
the hierarchy of the DNS.  Each zone has at least one ``authoritative'' DNS
server while one server can be authoritative for multiple zones.  In addition,
a nameserver can also be queried by client hosts and provide name
resolution for arbitrary domains for which it is not necessarily authoritative. Most
non-authoritative DNS servers only serve clients within their administrative
domain, e.g., an Internet Service Provider (ISP) providing name resolution for
its customers. However, services like OpenDNS and the Google public DNS provide
public available name servers.  

While some services are intentionally open to the public there is a large
mass of \mked servers unintentionally providing public name
resolution. DNS by default uses connectionless UDP and
usually the responses are larger than the queries as the query is contained in
the response.  Thus, DNS, by design, can be abused for amplification attacks,
in particular both with open resolvers~\cite{rossow2014amplification} and
resolvers that return large answers.  


Mitigation strategies against amplification attacks exist and are
usually deployed by the large providers of open DNS servers. Nevertheless, this
kind of abuse can not be completely prevented due to inherent protocol
limitations~\cite{rossow2014amplification}. While DNSSEC is currently being discussed as
mitigation for various other problems in the DNS protocol suite it exacerbates
this abuse as it often produces very large answers~\cite{dnssec}.

Another attack vector is information disclosure in reverse DNS lookups. Using
these an attacker may infer which hosts offer which services,
even inside a firewalled network, or disclose the organizational structure. Some
\mked systems may still allow zone transfers of full DNS
Zones~\cite{kalafut2008understanding}, which was historically the
default~\cite{djb}. 

Currently, we find more than 10,000,000 DNS servers on the Internet. A
substantial fraction, more than 5,000~\cite{streibelt2013exploring}, of these are
open DNS servers of which it is unclear to what extent they deploy
even the available limited amplification mitigation.

\smallskip
\noindent{\bf{SMTP:}}
\label{sec:smtp}
The objective of the Simple Mail Transfer Protocol (SMTP) as documented in 1982
in RFC821~\cite{rfc821} and updated by RFC5231~\cite{rfc5231} is to reliably
and efficiently transfer email.  Among the important features of SMTP is the
ability to relay email across multiple networks.

The base architecture of SMTP used open relays for forwarding messages between
Mail Transfer Agents (MTAs) without authentication or authorization, see
RFC822~\cite{rfc822}. Authentication was suggested by an Internet draft in
1995 and added with RFC2554~\cite{rfc2554} in 1999. TLS was also added
in 1999 with RFC2487~\cite{rfc2487}.

Attackers realized early on, that open relays are great for amplifying the
effects of worms, viruses, and in particular
SPAM~\cite{rfc1869}. Even with the CIA features SPAM is a
major daily annoyance.

These problems are usually mitigated by providing strict and well documented
default configurations~\cite{jung2004empirical}. In today's deployments almost
all SMTP servers only accept emails for their configured domains. Thus, the
possibility of amplification has been reduced.  If MTA to MTA relay is
allowed it is only with credentials and TLS.

Another problem with SMTP is that an attacker may take over an SMTP
server and send
rogue data which is not easily mitigated. The defense here is blacklisting,
whitelisting, sender verification, etc. see, e.g.,~\cite{cormack2007email}.

However, it is still possible - in an attempt to ``Make Things Work'' - to
misconfigure SMTP servers. After all, problematic configurations do still have
applicable use-cases on the Internet, e.g., an outbound email relay for a large
network which every machine should use.  In the wild, open SMTP relays are still
observed from time to time. But most are quickly found and closed down.

\comment{

braindump florian:
 - still often no real authentication
 - if IP od sender is in /24, accept email, because the net os mine
 - problem in webapp -> spam via webapp and local smtp relay

 - MX the same: when I can take over an IP of a mailserver I receive all mail
   for all domains the server is MX for, DANE is the first mitigation here,
   maybe, if all senders(!) verify the entries in DNS -> dnssec!

 - some people use MX on dialup with chaning IPs, other user gets IP, gets all mail

}

\comment{
\begin{itemize}
    \item Protocol to transfer mail between hosts
    \item First defined in 1982, RFC 821  -  most recent RFC 5231 (extended SMTP, used today)
    \item Base idea of the architecture were open relays which forward messages (see RFC822) without considering authentication/previliges
\end{itemize}

\begin{itemize}
    \item Initially designed without any authentication/authorization/tls
    \item Authentication added with RFC2554 in '99 (first draft in '95)
    \item TLS added in 1999 rfc 2487
\end{itemize}

\begin{itemize}
    \item Arising issue: Open relays were used to amplify worms, viruses, spam
    \item Even with extended SMTP the underlying issues from the 90s prevail
\end{itemize}

\begin{itemize}
	\item{documentation/defaults not wrong}
\end{itemize}

\begin{itemize}
	\item{issue known in academia jung2004empirical}
	\item{also known elsewhere  -  1995 rfc1869 calling issues endemic}
	\item{claimed to be mitigated jung2004empirical}
	\item main issue: owned servers sending rogue data  -  not really mitigatable.
    \item approached by (a) SPF and DKIM that add pointers who may send a mail
      from a domain to dns/sign mails from correct senders (b) wide use of 
      DNS blacklists, servers using them do not accept mail from servers that
      are open relays.
\end{itemize}

\begin{itemize}
    \item \mk as open relay is still seen in the wild from time to time in
      single systems  -  mostly mitigated due to DNS blacklists jung2004empirical
\end{itemize}

As already mentioned, the most common \mk issue with SMTP is that of an open
relay. An open relay is a SMTP server that accepts mail to any destination from
any user. This is an issue in the context of spam. With the creation of spam~\cite{}
best practices were established that exclude specific types of hosts from delivering
mail to mail exchange servers. This usually relates to hosts behind dynamic 
address ranges~\cite{}, which are also those usually making up the largest share
of botnets~\cite{}. This can be efficiently masked by abusing an open relay.
Similarly, this technique also can be used to efficiently hide the identity of
the spammer, a feature most crucial if the spam message are, for example, 
used for phishing.

Furthermore open relays allow the amplification of send mail messages. A single
mail message can have a multitude of blind carbon copy recipients. By using an
open relay an attacker only has to send one single mail from a possibly low
bandwidth connection, which is then re-distributed by the high bandwidth 
mailserver acting as open relay. 

These issues are usually approached by implementing strict default configurations.
However, it is still possible  -  in an attempt to ``Make Things Work''  -  to
misconfigure SMTP servers. This is mostly related to the fact, that the  - 
on the Internet  -  problematic configurations do still have applicable use-cases.
An example is an outbound mail relay for a large network, via which every machine
from that network should relay.
}

\label{sec:rip}

\smallskip
\noindent{\bf{NIS:}}
\label{sec:nis}
NIS has been initially developed by SUN~\cite{} to have a method that ensures
user identities, authentication and authorization information is present on
all machines under one authorative domain. The service is, by design, client
side unauthenticated, exposes all user passwords in maximum 3DES encryption 
to the world and does not support transport layer encryption.

\emph{Attacks:}
The attacks utilizing NIS are obvious. If it is exposed on the Internet a
remote attacker can dump the full user database, including the DES encrypted
passwords. Reversing these is not an issue for todays hardware~\cite{}.
Furthermore, the attacker obtains full information on all present users.

This issue can only be addressed by appropriate firewalling --- or replacing
NIS with a more steady method, e.g., Kerberos. The presented issues are also
not reduced by documentation indicating, that running NIS on another port
is an important security feature. Still, it can be held for these howtos, 
that they additionally recommend iptables rules.

\subsection{Discussion---\early}

A common \mk trap in this class is the assumption that neither client
authentication nor encryption is needed as the services are in a trustworthy
environment. Indeed, access without authentication is considered a feature
(FTP, SMTP, and DNS). Other abuses of \mk are stateless amplifications attacks
(TFTP, DNS) where the root cause is that the server sends data without checking
if the client wants it.

Based on these observations one would presume that protocols in this class have
seen the end of their live cycle. However, almost all of the above protocols
are still very popular.  The reason is (a) the Internet relies on the services
(DNS, SMTP) (b) their convenience (FTP), (c) the fact that there is no good
alternative (TFTP), and (d) the service happens to be running and is a legacy
service. Worse, the implementation of, e.g., TFTP requires a small code base which
makes it common in millions of customer premise equipment (CPE).
Indeed, these protocols are unlikely to disappear as they are the 
foundation of the Internet. 



The reason why such services are still in operation is twofold: Either it
is presumed possible to hide the services behind firewalls (TFTP, NIS, RIP) or
work on alternative protocols has started, but these protocols are not yet in
Internet-wide use (DNS, SMTP).  But, \mks occur if either the firewall fails or
the protocols are not used as originally designed.

The first major security incident which exploited a \mk was an email
amplification attack - namely the Morris worm in
1988~\cite{spafford1989internet}. At about the same time, the community started
to realize that major \mks can occur, see RFC1222~\cite{rfc1222} and
RFC1223~\cite{rfc1223} that ``provide guidance for vendors, implementors, and
users of Internet communication software''.


\section{Emerging Threats}
\label{sec:c2}
\setcounter{paragraph}{0}

Security incidents such as the Morris worm changed the way that Internet
protocols are perceived. Instead of designing them for the Internet at large,
they became explicitly designed with firewalls in mind. Thus, network firewalls,
more precisely packet filters~\cite{chapman1992network}, became the typical way
of fencing off network services.

Bellovin and Cheswick~\cite{bellovin1994network} state that the motivation for
Network Firewalls is: \emph{``Computer security is a hard problem. Security on
networked computers is much harder. Firewalls (barriers between two
networks), when used properly, can provide a significant increase in computer
security.''}  In addition, the approach from 1988 onwards is, according to
Bellovin and Cheswick~\cite{bellovin1994network}: \emph{``Everything is guilty
until proven innocent.  Thus, we configure our firewalls to reject
everything, unless we have explicitly made the choice - and accepted the risk
- to permit it''}.

Protocol designers find network firewalls to be a convenient way to handle
security. In their minds, firewalls enable them to basically ignore security
threats as they presume that the firewall rejects everything that is
``untrusted''. The design assumption of most protocols is that since the
attacker is not strong enough to get past the firewall the protocol itself can
be designed for a trusted environment.

However, as stated by Wool~\cite{wool2004quantitative}: \emph{``The protection
  that firewalls provide is only as good as the policy they are configured to
  implement. Analysis of real configuration data shows that corporate firewalls
  are often enforcing rule sets that violate well established security
  guidelines.''} His conclusion is to keep firewall configurations simple
but efficient to avoid \mk.

Thus, the threat model for this class is: ``weak attacker'' with
\perfect.  The representative protocols we take a closer look
at are: NetFlow, DHCP, and, iSCSI. Other protocols in this
class include: SNMPv2, Munin, NFSv3, Wake on Lan, Remote DMA, NBD, rsyslog, SCADA, early versions of
CIFS/SMB, and Mapping of Airline Traffic over Internet Protocol (MATIP).

\subsection{Example Protocols}
\setcounter{paragraph}{0}

\smallskip
\noindent{\bf{NetFlow:}}
\label{psi:netflow} 
Cisco Systems NetFlow service allows network administrators to collect IP flow
information from their network. A flow is a summary of a set of packets
that pass through a device that have some common property. NetFlow uses UDP as
its transport protocol. NetFlow is widely used in many ISP and enterprise
networks. Indeed, many resource accounting as well as security incident systems
are built on this data source. Early versions are documented as Cisco white
papers. Version~9 is documented in RFC3954~\cite{rfc3954}. 

NetFlow itself does not provide authentication, authorization, or encryption
support.  This was a conscious choice by the protocol designers, to cite
RFC3954~\cite{rfc3954}: \emph{``The designers of NetFlow Version 9 did not impose any
confidentiality, integrity or authentication requirements on the protocol
because this reduced the efficiency of the implementation and it was believed
at the time that the majority of deployments would confine the Flow Records to
private networks, with the Collector(s) and Exporter(s) in close proximity.''}
Indeed, RFC3954 specifically redirects the issue of security to the subsequent 
IPFix security requirements in RFC3917~\cite{rfc3917}.

RFC3954 outlines possible attacks including disclosure of flow
information data, forgery of flow records or template records, and DoS attacks
on NetFlow collectors.  The latter enables an attacker to exceed the
collector's storage or computational capacity and, thus, can disable the
monitoring of the network.
Using forgery, an attacker can inject flow information that (a) may redirect
network-forensic investigations by incriminates another party or (b) lead to
wrongful charges if NetFlow is the basis of accounting.

These attacks can, in principle, be mitigated by, e.g., moving to TCP and
enforcing TLS/DTLS and
mutual authentication. An example of such a mitigation strategy is the proposal
in the IPFIX security requirements RFC3917~\cite{rfc3917}, see
Section\ref{sec:c3}.  

The documentation of NetFlow is given mainly by Cisco White papers and Cisco
device configuration examples. We observe that the documentation does not even
mention how to secure NetFlow or that it is necessary. It does, however, point
out that NetFlow data can be used as security enhancement to investigate
network anomalies.

In summary, NetFlows main \mk trap is insufficient fencing. 
Since NetFlow is a stateless write only protocol it is infeasible to
estimate the \mked number of NetFlow collectors by active scans.

\smallskip
\noindent{\bf{DHCPv4:}}
\label{sec:dhcp}
The Dynamic Host Configuration Protocol (DHCP) is a stateful client-server
protocol that can be used to provide configuration parameters to hosts - the
clients - connected to the Internet. In practice, it is often used by clients to
retrieve their IP address configuration as well as additional parameters
including nameservers, domainnames, or local TFTP servers for diskless clients.

DHCP is based on the Bootstrap Protocol (BOOTP) and is documented in
RFC1531~\cite{rfc1531} in 1993. DHCPv6, standardized in RFC3316~\cite{rfc3315}
in 2003, tackles many of the security issues of DHCPv4.
Since DHCPv4 is still commonly used for configuring
IPv4 networks we discuss it in this section. In the following when we refer to
DHCP we mean DHCP for IPv4 address configuration.

Regarding security RFC1531~\cite{rfc1531} claims that ``\emph{ DHCP is built directly on UDP
and IP which are as yet inherently insecure}''. Since DHCP does not add any
security features itself this means that the protocol
lacks all basic security features. It has been designed without
client authentication, server authentication, or encryption.
Client Authentication was added in 2001~\cite{rfc3118}, but is not widely 
implemented, especially in the common embedded DHCP servers for CPEs.

As a result, any attacker can exploit the  possibility for a single
client to request all leases held by a DHCP server. This effectively blocks all
other clients from obtaining an address. This attack is critical as it can be
executed locally as well as remotely if the DHCP server is accessible from the
Internet.

The next problem is that servers do not have to authenticate themself towards
the clients. Therefore, any host can pretend to be the authoritative DHCP server
for its network segment. This allows an attacker to impersonate a DHCP server
and send malicious information to the clients, e.g., to use (a) a different
gateway which is hijacked by a monkey in the middle or (b) a different DNS server
to spoof internal websites and access credentials.

Among the common DHCP servers are ISC-DHCP and dnsmasq. Their documentation is
reasonable but ignores the topic of security. 
The main \mk trap is yet again insufficient fencing. 
Indeed, during the 28th Chaos Communication Congress in December 2011 the 
network operations team observed a DoS against their publicly reachable 
DHCP server. A virtual machine hosted in Amazon EC2 performed a lease starvation
attack on that system~\cite{ccc2011}.

Since DHCP is the common protocol for assigning dynamic IP addresses it is
in common use almost everywhere. Indeed, even many home users use DHCP due to
the large scale introduction of Network Address Translation (NAT) enabled 
home routers.

\smallskip
\noindent{\bf{SNMPv2:}}
\label{psi:snmpv2}
The Simple Network Management Protocol (SNMP) dates back to
RFC1067~\cite{rfc1067} from 1988 and is the standard protocol for managing IP
network devices, including routers, switches, workstations. It is typically
shipped on the device. Small command extensions led to Community-based
SNMPv2~\cite{rfc1901} the version supported by most vendors, e.g., Cisco and
Juniper. While other versions of SNMPv2 already support extensive security
features most became prominent with SNMPv3~\cite{rfc2571,rfc3410}, which we
discuss in Section~\ref{sec:c3}. The only authentication mechanism of SNMPv2c
is the so called community string. Moreover, SNMPv2c does not support transport
layer security.

Common \mks are the use of weak (default) credentials, e.g., the community
string \verb+public+ for read access and string \verb+private+ for the
read/write access~\cite{moonen2012digitale}.  While SNMP enabled devices should
be shielded by proper firewall configurations they often are not. Moreover, SNMP proxies which are
designed to handle ACLs can easily be misconfigured as well. Note, weak
credentials can lead to disclosure of information. With the \verb+private+
community it is possible to take over the device. Moreover, open SNMPv2 servers
have been used for amplifications attacks~\cite{rossow2014amplification}.

By default, a lot of devices come with communities preconfigured,
e.g., \verb+public+ or/and \verb+private+. Even if the operator configures their
own communities they  often forget to remove the preconfigured ones - leaving
the door open to attackers. Moreover, with a network sniffer it is possible to
extract community strings. Depending on the class of the device, the documentation
differs significantly from good for high-end devices to almost none for
low-end customer premise devices. This is, in particular, problematic as the
customer premise devices are often directly connected to the Internet.  The
potential number of devices that may be subject to this class of
misconfiguration is, according to Shodan scans, more than 3,800,000 devices
with the \verb+public+ community string.

\comment{
\begin{itemize}
    \item weak credentials
    \item comes in large amount with default communities public and private
\end{itemize}

\begin{itemize}
    \item 3,800,000 devices that have public set as community string
    \item data from shodan 
\end{itemize}
}

\smallskip
\noindent{\bf{Munin:}}
\label{psi:munin}
Munin~\cite{munin} is an open source networked resource monitoring tool. It is a
simple service allowing retrieval of server statistics for monitoring
purposes.  The monitored server listens on an open port for inbound
connections.  The Munin monitor polls each of the targets by connecting to the
port and requesting the status data.
Other similar services use the same basic schema, e.g., NCSA, Ganglia, Collectd  etc.
Authentication is implemented by whitelisting  IP addresses or address
ranges of Munin servers. Authorization and transport security are not available.

Common \mks are weak firewalls together with too liberal ACLs. This
allows attackers to obtain detailed information about the infrastructure as
well as fine-grained usage information. This, in turn, can enable a whole range
of security critical side channel attacks on the cryptography of other
protocols~\cite{zhang2012cross}. Moreover, given recent side-channel attacks
that use acoustic signals from the CPU~\cite{genkin2014rsa}, it is not
unlikely that attackers can use such data to extract, e.g., secret keys
from the monitored servers.

While the documentation states that ACLs have to be clearly limited to
authorized hosts, we still find more than 6,000 systems in the Shodan data.

\smallskip
\noindent{\bf{NFSv3:}}
\label{sec:nfsv3}
The Network File System (NFSv3) offers transparent access to remote files. It
has become one of the common UNIX network file systems. It is documented in
RFC1813~\cite{rfc1813} from 1995. NFSv4 or more general NFS with Kerberos
support is discussed in Section~\ref{sec:c3}.

NFSv3 offers host-based authentication on network-wide names but not per
principal authentication. Moreover, NFSv3 relies on the client OS for
authorization. On the wire encryption is not supported~\cite{rfc1813}.  While
NFSv3 in principle supports Kerberos secret keys it does not mandate them and most
deployments do not use them.  System lore warns to use NFS without Kerberos, see
Section~\ref{sec:c4}, if strong security is needed, e.g., ``Kerberos is key for
secure data access and not NFSv4''~\cite{ibm-nfs}.

Common \mks for NFSv3 involve the Access Control Lists: They can be too
liberal, e.g., network wide, or incorrectly specified, e.g., 
wrong subtree of the file system.  In either case, an attacker can mount an NFS
share and read or modify arbitrary files.
While the available documentation stresses the importance of ACLs, \mked servers can be found in the wild.
These problems are widely known both in systems lore~\cite{rfc2623} as well as
academia~\cite{tanenbaum2007distributed}.

\smallskip
\noindent{\bf{iSCSI:}}
\label{sec:iscsi}
The Internet Small Computer System Interface (iSCSI) is a protocol for remotely
accessing block devices over the Internet using SCSI commands first documented
2004 in RFC3720~\cite{rfc3720}. Since iSCSI was designed for a hostile
Internet, a dedicated RFC~\cite{rfc3723} exists, that spells out the security
requirements for iSCSI and similar network accessible block storage
protocols. This RFC has been updated most recently in April 2014 by
RFC7146~\cite{rfc7146}.

These RFCs require iSCSI to include authentication and authorization but
delegate encryption and integrity to IPSec. Moreover, RFC3723~\cite{rfc3723}
acknowledges common threats for iSCSI deployments under the assumption that
authentication and authorization are working, i.e., that the attacker is not
able to initiate a valid connection.
Moreover, the base security assumption is that there is no monkey in the middle
either due to IPSec or an isolated network segment.

However, if due to a \mk the iSCSI target is reachable via the Internet without
authentification iSCSI becomes a severe security liability.  An attacker
with access to an iSCSI volume can tamper with all data thereon and can take
over all machines with root file systems on those volumes.

Indeed, it is easily possible - and many large enterprise applications and
howtos for setting up UNIX based targets recommend - to configure iSCSI without
any authentication, neither for the whole target set nor the individual
targets.  Usually, this is done to cater to operational needs, especially in
the cases where iSCSI volumes are used as boot devices or if a dynamic set of
virtual machines has to have access to a volume.  In contrast to the above
common malpractice, the Payment Card Industry Data Security Standards
(PCI-DSS)~\cite{dss2014payment} explicitly requires client authentication, in
particular, it requires it for each volume individually.  Nevertheless, the
common \mk trap for iSCSI is missing authentication coupled with reliance on
fencing.

So far, these security issues have been recognized in the
industry~\cite{dwivedi2005iscsi} but not necessarily in academic references.
With a quick zMap scan, we find roughly 9,000 iSCSI targets reachable via the
Internet of which 1,000 do not require authentication.  Among these are various
major organizations as well as academic institutions.

\subsection{Discussion---\emerging}

The base assumption of all protocols/services in this class is that the local
LAN is safe. Thus, a common misconception is that they can be used with
``convenient'' security settings. This often leads to major security incidents
when the fencing mechanisms fail. This is the major \mk trap for all protocols
in this class. 

NetFlow is a blatant example of the low security considerations within this
class. Its security concept  relies entirely on fencing. DHCP goes even further:
first anyone can run a rogue DHCP server and second engineers think about DHCP
as a link layer protocol. SNMPv2 is one of the protocols in this class that
first added security but then reduced it. NFSv3 does do authorization using
OS ACLs, but without authentication. This means that anyone can impersonate
anyone.  While iSCSI, in principle, supports authorization and authentication
some deployments do not enable it as iSCSI should be restricted to the storage
network and, if used as boot device, is difficult to supply the clients with
secured credentials for the iSCSI volume. However, if fencing breaks down this
is a major \mk as large amounts of sensible data are leaked.

However, the assumption that everything can be fenced in does not necessarily
hold as specifying security policies is difficult and realizing them in
a firewall is rather difficult and prone to
errors~\cite{mayer2000fang,al2003firewall,cuppens2005detection,yuan2006fireman,garcia2013management}.
Among the complications are that the designer of the security policies
are not necessarily the ones that configure the firewalls and those are not
necessarily the ones that deploy the network services. Moreover, updating and
maintaining such rules is quite error prone. This opens up the network
service for all kinds of attacks that bypass firewalls or access services that
are thought to not be reachable from the Internet.

Other means of fencing include: (a) not connecting the service to the Internet
at all (air gap) (b) VLANs and sub-networks (c) Virtual routing and forwarding
(VRF).  But, there are known attacks to all of them. Examples of how firewalls
are circumvented via VPNs, hidden dialups/UMTS and other covert channels are
described in the Maroochy water breach
discussion~\cite{slay2008lessons}. Stuxnet~\cite{falliere2011w32} is the prime
example for bypassing an air gap.  One example for broken VRF is accidentally
announcing a BGP full table into the VRF engine.  Overall, the industrial lore
states, from the Security Issues and Best Practices for Water/Wastewater
Facilities~\cite{hayes2013security}: \emph{``Industrial networks are often
  shared with the business side of the operation. VLANs, sub-networks,
  firewalls all help to create a layer defense, but are not impervious.''}.

This is  particularly the case for services that were first envisioned for
enterprises and  then commonly used in Small Office/Home Office (SoHos)
and home networks. In these settings security by default configurations are
essential as the users often lack the knowledge and means to properly address
security and network challenges.

\section{Complex Security Solutions}
\label{sec:c3}
\setcounter{paragraph}{0}

At the end of the 20th century the awareness that firewalls were not ``the''
security solution became prominent. For example, RFC3365~\cite{rfc3365} dated
2002 states: \emph{``History has shown that applications that operate using the
  TCP/IP Protocol Suite wind up being used over the Internet.  This is true
  even when the original application was not envisioned to be used in a ``wide
  area'' Internet environment.  If an application isn't designed to provide
  security, users of the application discover that they are vulnerable to
  attack.''}

As a result, protocol designers realized that (a) there was a need for improved
protection architectures, e.g., the work on SANE~\cite{casado2006sane} or
DoS-limiting network architecture~\cite{yang2005limiting}, and (b) that
security had to be an essential feature of future protocols and services.
Moreover, just adding another component to ensure security to existing
protocols, e.g., firewalls, did not suffice.  This fits the increased need for
security in the society  due to the increasing economic relevance of the
Internet~\cite{mahadevan2000business}.

At the same time the diversity of the scenarios also increased with home users,
SoHos, enterprises, infrastructure providers, company mergers and splits,
etc. Indeed, road worriers started to appear.  As a result, 
more assets were at stake which had to be accessible in many different ways. Thus,
versatile security solutions to model complex organizational
structures, e.g., via role-based access control (RBAC), were needed.

Indeed, the Danvers Doctrine~\cite{rfc3365} stated that the \emph{``IETF should
standardize on the use of the best security available''}. Thus, the threat
model for this class is: ``strong attacker'' with ``\perfect''.  The representative
protocols we take a closer look at are: IPP, SNMPv3, and IPFIX.  Other
protocols in this class include: LDAP-ACL, NFSv4, AFS, Postgresq, FTPs, RADIUS/WPA2Enterprise, s/MIME encryption,
SSL/TLS, PGP, and seLinux, as an example from system security.

\subsection{Example Protocols---\complex}
\setcounter{paragraph}{0}

\smallskip
\noindent{\bf{IPSec:}}
\label{sec:ipsec}
IPSec, first introduced in RFC1825--1829~\cite{rfc1825} and updated by
RFC4301--4309~\cite{rfc4301}, is a suite of protocols that promise to
seamlessly extend IP with authentication, data integrity, confidentiality,
non-repudiation, and protection against replay attacks.

To cite Ferguson and Schneier~\cite{ferguson2000cryptographic}: \emph{``Our
main criticism of IPsec is its complexity. IPsec contains too many options and
too much flexibility; there are often several ways of doing the same or similar
things. This is a typical committee effect.''} Thus, this is a prime example of
this class.  However, IPSec is often used as an argument why it is possible to
leave out certain security features in other protocols~\cite{rfc3723,rfc4297}.

IPSec is in use for corner cases such as LTE backend network
security~\cite{bikos2013lte}. Indeed, as Ferguson and Schneier
state~\cite{ferguson2000cryptographic}: \emph{``Even with all the serious
criticisms that we have on IPsec, it is probably the best IP security protocol
available at the moment.''}  Still, IPSec has not yet seen widespread
deployment, e.g.,~\cite{richter2015distilling}. One possible reason is usability. Here we
cite Gutmann~\cite{gutmann2005security} \emph{``If we consider security
usability at all, we place it firmly in second place, and anyone wishing to
dispute this claim is invited to try setting up an IPsec tunnel via a firewall
or securing their email with S/MIME.''}

\smallskip
\noindent{\bf{LDAP:}}
\label{sec:ldap}
LDAP, the Lightweight Directory Access Protocol, is a protocol for accessing
and maintaining distributed directory information. It builds upon the ideas of
X.500 but differs, in particular, with regards to security
features~\cite{hassler1999x} and simplicity. LDAP is designed to be extensible and
flexible, see the many LDAP related RFCs, 
including RFC2251~\cite{rfc2251} dated 1997, and RFC4510~\cite{rfc4510} to
RFC4519~\cite{rfc4519}.  LDAP organizes its data in a tree like
ASN.1~\cite{asn1}.
It is often used to organize organizational information, groups, and,
in particular, user data, including account information, personal
information, and authentication data.

LDAP offers transport security. It also provides various forms of
authentication, including SASL and Kerberos, see RFC4513~\cite{rfc4513}. 
The authorization concept of LDAP is probably one of the most complex, yet,
also most powerful systems currently available. In LDAP, the ACL
roughly follows role-based access control~\cite{rfc4513}. It distinguishes between
anonymous, authenticated, and specific connections. Specific connections
are defined by properties on the object that holds the connection. This may
be, but is not limited to, group membership, subtree membership, tree position,
attributes in the object, etc.

The main misconfiguration opportunities for LDAP are (a) that operators use LDAP
over the Internet without transport layer encryption and (b) that operators make
mistakes while setting up access
control~\cite{xu2013not,findlay2011best}. 

Especially when used for authentication, the bootstrap process is hard.
Clients have to be configured and authenticated correctly, as well as
authorized to access the information in the LDAP tree, and use it to perform
authentication and authorization within their own applications.  Furthermore,
no reasonable default values can be created for ACLs, as these depend on the
root nodes of the LDAP tree, which is organization dependent.  To cite
RFC4513~\cite{rfc4513}: \emph{`` Operational experience shows that clients can
(and frequently do) misuse the unauthenticated authentication mechanism of the
simple Bind method see (Unauthenticated Authentication Mechanism of Simple
Bind)``} 

Commercial distributions come with reasonable pre-configured ACLs. The
non-commercial ones usually come with one rule, no write access for users that
are not system administrators. 

\smallskip
\noindent{\bf{IPP:}}
\label{sec:ipp}
The Internet Printing Protocol (IPP) is documented in
RFC2565~\cite{rfc2565} dated 1999 and its companion RFCs. IPP is an
application level protocol suite for distributed printing using Internet tools
and technology. It uses HTTP, namely 1.0 or 1.1 as its transport protocol.

IPP itself implements the relevant mechanisms to perform strong
authentication - by default against members of a local UNIX group via PAM
(Pluggable Authentication Module) and supports the use of transport encryption.
For IPP~\cite{rfc2565} authentication and authorization are critical due to an
unrelated security topic, accounting.  Printing, or rather, use of paper and
ink, must be accounted for in most companies as well as universities.  Thus, it
is not surprising that the most prominent UNIX based IPP server, CUPS, currently
uses TLS for all connections containing credentials. Moreover,
authentication is required by default for all administrative actions.

The main \mk trap with IPP is that printing on a device is - by default -
allowed for unauthenticated clients~\cite{printerinsecurity}. Thus, a remote
attacker can print on all printers they learn about. While this is usually not
a major security problem, it may become an interesting basis for social
engineering attacks. In addition, it enables DoS attacks, e.g., if an attacker
prints endless numbers of fully black pages. Lastly, it is a nasty way of large
scale resource waste. Keep in mind that the IPP service is offered by most major
network attached printers by default as well as apple devices if they share
their home printer.

\smallskip
\noindent{\bf{NFSv4 with Kerberos:}}
\label{sec:nfsv4}
NFSv4 is another distributed file system protocol. Unlike its predecessor NFSv3,
see Section~\ref{sec:c2}, NFSv4 has support for strong security and its
negotiation built in. NFSv4 uses a principal-based authentication model 
rather than machine-based as prior versions of NFS did.

The NFS standard~\cite{rfc7530} mandates strong security using
Kerberos~\cite{rfc4121}. Kerberos, according to Neuman et
al.~\cite{neuman1994kerberos}, is a distributed authentication service that
allows processes of a principle to prove their identity to an application server
without sending data across the network that might allow an attacker to
impersonate the principal. Kerberos also provides integrity and
confidentiality.

This sounds great.  However, while the Storage Networking Industry Association
(SNIA)
states~\cite{snia2015nfs} that \emph{``With careful planning, migration to NFSv4.1 and NFSv4.2 from prior
versions can be accomplished without modification to applications or the
supporting operational infrastructure, for a wide range of applications; home
directories, HPC storage servers, backup jobs and so on.''}  system lore says
that NFSv4 deployment is lacking.

Indeed, NFSv4 may come with a performance penalty~\cite{chen2015newer} and, as
McDonald points out \emph{``One area of great confusion is that
many believe that NFSv4 requires the use of strong security. The NFSv4
specification simply states that implementation of strong RPC security by
servers and clients is mandatory, not the use of strong RPC security. This
misunderstanding may explain the reluctance of users to migrate to NFSv4, due
to the additional work in implementing or modifying their existing Kerberos
security.''}

Thus, we conclude that even though strong security options exist, often
system administrators choose to not deploy them. One reason is the
inherent complexity of Kerberos. Indeed, Bouillon in his Black Hat EU 2009 talk
stated~\cite{bouillon2009taming} \emph{``However, lots of system administrators still
make dramatic mistakes while configuring it (those mistakes are made more
likely by buggy GUIs and their poor documentation)...''}. Indeed, from
discussions with several operators we learned that they often hesitate to
deploy Kerberos due to its complexity, lack of documentation, and difficulty 
debugging its deployment. 

As a consequence it is not surprising that the Shodan scans find many NFSv3
deployments (100,000) but no NFSv4 deployments. Even though NFSv4 should be
easier to detect as it uses the IANA port 2049 and does not rely on portmap.

\smallskip
\noindent{\bf{SNMPv3:}}
\label{sec:snmpv3}
SNMPv3~\cite{stallings1998snmpv3,rfc3410} is the successor to the Simple Network
Management Protocol~v2, see Section~\ref{sec:c2}. The motivation for SNMPv3,
RFC3410~\cite{rfc3410} was to fix: \emph{``The unmet goals included provision
of security and administration delivering so-called ``commercial grade''
security with: authentication \ldots, privacy \ldots; authorization and access
control; and suitable remote configuration and administration capabilities for
these features.''}  Thus, SNMPv3 makes few changes to the protocol aside from
adding the option of on-the-wire encryption. Rather, it focuses on two main
aspects, security and administration~\cite{corrente2004security}.
SNMPv3 supports the notion of users and authorization.

The main \mk trap with SNMPv3 is using SNMPv2 instead or in parallel. Indeed,
SNMPv2 and SNMPv3 are not exclusive. A host offering SNMPv2 can also offer
SNMPv3 even on the same port. Today many hosts indeed offer SNMPv2 and SNMPv3
simultaneously~\cite{rfc2576,rfc3584}.  Since SNMPv2 is often enabled by
default the hosts remain vulnerable even though they support SNMPv3. Therefore,
it is not surprising that we still see more than 3,800,000 hosts with SNMPv2
enabled. Indeed, this is supported by data from Cisco Advanced Services from
May 2013~\cite{cisco2015snmp} which reports on the SNMP configuration status of
1,724,827 device configurations in 2013, see Table~\ref{tbl:c}.
\begin{table}
\begin{center}
\caption{Adaption of SNMPv3 over time following~\cite{cisco2015snmp}.}
\label{tbl:c}
\begin{tabular}{|c|r|r|r|}
\hline
  protocol & customer & devices & change from\\
  version  &          &         & 2012 -- 2013\\
 \hline
 \hline
 SNMPv2 & 98.5\% &  88.2\% & up 9\% \\
 SNMPv3 & 34.6\% &  10.4\% & up 4\% \\
\hline
\end{tabular}
\end{center}
\vspace{-0.8cm}
\end{table}
Even though the adoption of SNMPv3 has increased it has been at a lower
rate than SNMPv2. Part of the reason is the complexity of SNMPv3 as outlined,
e.g., by Cisco training material~\cite{cisco2015snmp}.

However, even if SNMPv3 is correctly deployed, the corresponding RFCs come with
another \mk trap. The original RFCs, RFC2264~\cite{rfc2264} to RFC3414~\cite{rfc3414},
only specify DES as a cipher.  AES was added significantly later, RFC3826~\cite{rfc3826}.
Similarly, the only supported digest algorithms are MD5 and SHA1. This 
leads to implementations with weak crypto and, thus, 
are susceptible to advanced attacks~\cite{lawrence2012under}. 
SSHv1 suffers from similar problems as CRC32 is fixed in the protocol
specification~\cite{barrett2005ssh}.

\smallskip
\noindent{\bf{IPFix:}}
\label{psi:ipfix}
The purpose of the IP Flow Information Export (IPFIX) protocol, see
RFC5101~\cite{rfc5101} from 2008, is to transfer IP Traffic Flow information
from an exporter to a collector. IPFIX is the vendor-independent successor of
NetFlow version~9, see Section~\ref{sec:c2}.  IPFIX supports flexible
definitions of network flows via a template based extensible information
model. The design goal was to make the protocol future proof as well as
applicable to all network protocols.

Given the inherent insecurity of NetFlow the goal of IPFIX was to incorporate
strong security in its design, see RFC3917~\cite{rfc3917}. This resulted in
RFC5101~\cite{rfc5101} which states that IPFIX must ensure confidentiality and
integrity of the transferred IPFIX data and authentication for the exporter and
collector. 

However, none of the major vendors, including Alcatel, Cisco,
and Juniper implement any of the CIA mechanisms of the IPFIX RFC. Thus,
the same exploits and \mks that apply to NetFlow also apply to
IPFIX, see Section~\ref{sec:c2}. Hence, the major \mk trap is
insufficient fencing.

\subsection{Discussion---\complex}

The common design of all protocols within this class is that they are designed
according to the Danvers Doctrine~\cite{rfc3365}. Therefore, they all offer full CIA support.
Unfortunately, when looking at the deployment base, we either find few indications
of actual use of the protocol or that the deployed instances do not utilize or implement
the full CIA support.

Among the reasons are difficult setup (LDAP, NFSv4) or limited
perceived benefit of latest version of the protocol (NFSv4, SNMPv3).  Also, they
are often used within the SoHo, rather than the enterprise, where there is a lack
of required security infrastructure (IPP without Kerberos). Furthermore,
implementations do not support required protocol security features
(IPFIX). Furthermore, third party documentation may recommend simpler
solutions, e.g., Postgres.

Another aspect is that the system administrator and the IT security
professionals are typically not the same person,
e.g.,~\cite{botta2007towards,furnell2009integrated}.  Moreover, these systems
come with substantial complexity which leads to many \mk traps. For example,
Xu et al.~\cite{xu2013not,xu2015hey} in their papers \emph{``Do Not Blame Users
  for Misconfigurations''} and \emph{``Hey, You Have Given Me Too Many Knobs!
  Understanding and Dealing with Over-Designed Configuration in System
  Software''} point out that major causes of today’s system failures are
\mks due to complexity. However, it is not necessarily the user or
the system administrator who is to blame, but rather the inherent complexity
and mismatch of the tools~\cite{haber2007design,barrett2004field} as well as
poor usability for system operators~\cite{xu2013not,xu2015hey}.

Given the above complexity, let's review what might or might not
motivate a system operator to deploy the latest secure protocol suites,
see, e.g., the discussion of West as well as Scheier in their papers on
\emph{``The psychology of
  security''}~\cite{west2008psychology,schneier2007psychology}: (a) No
or little reward for secure behavior. Indeed, there are hardly any
monetary incentives for deploying the secure versions. Some claim that
such incentives might change this~\cite{greenwald2004user}.  (b) The
misconception of operators that there is a low risk of being attacked.
(c) The ``laziness'' of the operator that wants to get the main service
working first and then worry about security if there is time left.  (d)
The time pressure by management that forces the operator to get a
service working in no time.  (e) The presumption that no one is
likely to get caught for not deploying the secure version of the
service. (f) The maintainability of the complex security infrastructure.

\section{A new Simplicity}
\label{sec:c4}
\setcounter{paragraph}{0}

The inherent complexity of protocol suites of the previous class provided
the motivation for trade-off based security.  This class is
about ``strong attackers'' and ``\good'' security.  
As noted by Bruce Schneier~\cite{schneier2007psychology}: \emph{Security is a
  trade-off. This is something I have written about extensively, and is a
  notion critical to understanding the psychology of security. There's no such
  thing as absolute security, and any gain in security always involves some
  sort of trade-off.}  

The protocol designs in this class must be seen in the context that we now
have clouds as well as many start-up companies with various (mobile)
applications. Indeed, the Internet is experiencing yet another growth
explosion in terms of services~\cite{armbrust2010view,pallis2010cloud}.  In
terms of infrastructure, we are now in a ``world of HTTP everywhere'',
virtualization, the cloud, large scale as well as microservices architecture,
and configuration orchestration.

Let us consider Internet application developers. Among the easiest ways
to develop new applications is to do it in the cloud using a
microservice architecture. They can rely on ``Node.js'' or
similar programming languages and reuse existing code. The reuse of
existing code has been made easy by the equivalent of appstores. As
transport protocol they will most likely use HTTP/HTTPs.
Moreover, the application grows as the feature sets and/or user
base increases. A common observation is that the security review of such
services is often lacking and the assumption is that it is possible to
fence the service in a private cloud as it is just another Web service.

With ``simplicity'' we refer to the security concept rather than the feature
sets of the protocols or, rather, protocol suites.  The threat model for
this class is: ``strong attacker'' with ``\good''.  The representative
protocols/protocol suites we examine are: Telnet, key-value
stores, VNC, and the many incarnations of (HTTP/Socket) APIs. Other protocols
in this class include: PulseAudio's and Systemd's internal protocols as well as
control protocols for tools such as Nessus, Aircrack, etc. Since most of these
use HTTP as transport layer protocols the same \mk traps apply as discussed in
the API subsection.

\subsection{Example Protocols}
\setcounter{paragraph}{0}
\smallskip
\noindent{\bf{Telnet:}}
\label{sec:telnet}
Back in 1969 the idea of the Telnet protocol, e.g., RFC15, RFC137, RFC854,
RFC5198~\cite{rfc15,rfc137,rfc854,rfc5198}, was to make a terminal usable by a
remote host as if it was local. The result was a bi-directional, byte-oriented
communication protocol. Since SSH~\cite{rfc4251} was designed, in 1995, as a
secure replacement for Telnet, rlogin, etc.~\cite{Ylonen1996} Telnet usage
should have declined to almost zero. Therefore, if at all we should have discussed
Telnet in Section~\ref{sec:c1}.

Unfortunately, today for many application and infrastructure systems, e.g.,
Customer Premise Equipment (CPEs), Storage Area Network (SAN) Devices, and what
is now the Internet of Things (IoT) and Industry 4.0, Telnet is again the
default choice for accessing devices~\cite{191966} - with attacks skyrocketing
in the past 18 months~\cite{191952}.  After all, Telnet is relatively
simple. Thus, most of these devices already come with a built-in daemon from
the original design/equipment manufacturer (ODM/OEM) that supports a subset of
the Telnet protocol in their firmware templates~\cite{costin2014large}. This is the reason
we discuss Telnet in this section.

The original versions of Telnet did not offer authentication/authorization or
encryption. Telnet authentication defaulted to the operating system's login.
Various
authentication and authorization mechanisms including Kerberos and RSA were
added to Telnet in 1993, see RFC1409~\cite{rfc1409}. Adding
TLS to Telnet was suggested by an Internet draft in
2000~\cite{draft-ietf-tn3270e-Telnet-tls-06}. A recent publication on 
vulnerabilities in telematic systems found that for all investigated devices
authentication was not enabled for the Telnet interface~\cite{191966}.
 
CPEs differ from operator equipment in the sense that they are actually
operated by the end users. They also differ from the typical end-user
equipment in the sense that they are often preconfigured and directly
reachable via the Internet.  Given vendors pre-provisioning their products
with known default credentials, these CPE devices can be vulnerable as soon as
they are accessible over the Internet. The extent to which this can be a
problem was demonstrated in 2012 by the Carna 
botnet~\cite{carna,krenc2014internet}. Another major attack used these devices
to change the DNS settings of the end-users and, thus, did  a monkey in the
middle attack~\cite{assolini2012tale}.

Mitigation is relatively easy. Vendors should rethink if Telnet access is
necessary at all. Even if it is, it should be possible to physically turn it
on/off with a small dip switch on the device - with default off.  Furthermore,
they should not come with no, default, or guessable credentials. Indeed, the
initial credentials should be unique for each device and should not be
computable from anything that is also related to the
device. Various vendors already use this approach.~\cite{191954}

The documentation of the Telnet service is usually very limited as the
documentation usually focuses on the devices themselves and only mentions that
Telnet is supported. In the past most CPE services used default configurations
with weak credentials. 

Today, we find more than 10,600,000 devices with an open Telnet port according to
Shodan. Indeed, the Carna botnet exploited about 1,200,000 of these devices~\cite{carna}. Furthermore, a recent study by Pa et al.~\cite{191952}
finds that the attack volume on Telnet enabled IoT devices has increased by
many orders of magnitude since 2014. 

\comment{
\begin{itemize}
    \item initially from 1968, for interconnecting terminals(rfc15)
    \item updated with RFC854 and 5198 ony to be superseeded by SSH
    \item even got support for TLS \cite{draft-ietf-tn3270e-Telnet-tls-05}
    \item BUT: This is not the Telnet we are talking about in this section!
	\item{orig 1969, rise of CPEs around 2002}
    \item new issue: SoCs that come with Telnet on their serial sun2002simulation horan2013serial guo2008hardware
    \item there: technically simple ascii-over-the-wire  -  not really Telnet
          compliant, but still on port 23 and usable with Telnet.
\end{itemize}

\begin{itemize}
    \item some have authn some don't
    \item cpe's usually single user systems
    \item tls usually not used
    \item main issue: No or weak default credentials. see defaults.
	\item have to authenticate well  -  well known e.g. rfc 1409
\end{itemize}

\begin{itemize}
    \item guessable/no credentials mean full access to device
    \item one can do \$everything. Example from the past: LARGE botnet
    \item but also: abuse to mitm/wrongDNS the endusers (cite brazil case)
\end{itemize}

\begin{itemize}
	\item spin: doc/defaults relates to delivered IoT/CPE devices
	\item usually: weak default credentials, no doc for running service
    \item there things start to look better. yet this issue circles
      around old CPE and all the cheap china plastic.
\end{itemize}

\begin{itemize}
	\item carna botnet used this krenc2014internet
    \item weak Telnet credentials age old issue - that there are so many devices
       again is new.
\end{itemize}

\begin{itemize}
    \item 10,600,000 with open p23 according to shodan
    \item carna found ~400,000 for its botnet.
    \item far to many
\end{itemize}

Telnet, as first defined in 1969's RFC15~\cite{} and updated by RFC854~\cite{} 
and RFC5198~\cite{}, is a simple, bi-directional communication protocol. It has 
famously been used for remote login to systems~\cite{}. Due to drawbacks in the 
Telnet protocol regarding encryption it has since then been superseded by Secure 
SHell (SSH). However, newer Telnet specifications also include Telnet with 
additional transport security~\cite{draft-ietf-tn3270e-Telnet-tls-05}.

\emph{Attacks:}
Telnet suffers from two major \mk or rather misconceptions. The first issue
is related to missing TLS. Used over an insecure  -  possibly wireless  -  
channel, credentials become known to attackers. The second issue relates to
weak credentials. While it is a common thing that users (and even administrators)
use weak credentials, this issue is  -  by now  -  mostly extinct in the Telnet
world. Most systems with large user bases have by now migrated away from Telnet.

However, for many application and infrastructure systems, e.g., CPEs, SAN
Storage Devices and what is now the IoT and Industry 4.0 Telnet is the default
choice for providing direct access to these devices. With vendors pre-provisioning
their products with known default credentials these devices are vulnerable as
soon as they are accessible from the Internet. The extent to which this proofs
to be a problem was demonstrated in 2012 by the creators of the carna botnet~\cite{}.

Mitigation is considerably easy. First, vendors should consider if Telnet access
is necessary at all. If it is, it should be possible to physically turn it on
with a small dip switch on the device  -  with a default-position of off. 
Furthermore, default credentials should be unique for each shipped device, and
not computed from anything about the device~\cite{}. Various vendors already
demonstrate, that such an approach is feasible~\cite{}.
}

\smallskip
\noindent{\bf{Key-Value Stores:}}
\label{sec:kv}
A useful Internet service is provided by key-value stores which are widely used by
companies such as Amazon, FaceBook, Digg, and
Twitter~\cite{atikoglu2012workload}.  Key-value stores provide a mapping
between keys and their associated data and can, if
implemented \emph{In-memory}, avoid classic I/O bottlenecks. Thus, they allow
quick non-relational data storage. Examples include Dynamo, MongoDB, Redis, and
memcached. Initial development started around 2004. A first
paper~\cite{decandia2007dynamo} reporting on ``Dynamo, Amazon's highly
available key-value store'' appeared in 2007.
While not a protocol in themselves key-value stores are an almost universal network
service and often rely on HTTP/HTTPs or JSON for their communication.

Initially, most of the key-value services did not support authentication,
authorization, and/or encryption. However, over time most were augmented with
support for authentication, authorization, and encryption.

To highlight the underlying assumption of the design of all of key-value stores we cite the Dynamo
paper~\cite{decandia2007dynamo}: \emph{``Other Assumptions: Dynamo is used only
by Amazon’s internal services. Its operation environment is assumed to be
non-hostile and there are no security related requirements such as
authentication and authorization.''}

Most key-value stores allow for transitive attacks. When an attacker can access
a central web session storage they can impersonate users and or
administrators. When using \kvs for caching, e.g., of SQL requests or of
pre-compiled Just-In-Time bytecode, access to the cache can result in
information disclosure or even attack code execution.

Next we give a few examples, using memcached, of the defaults when deploying
key-value stores. For memcached strong authentication on the basis of SASL was
introduced in 2009. Moreover, memcached now supports transport layer
encryption. However, this support is lacking by default in most
Unix-derivatives as the implementation of SASL was flaky in memcached and led
to bugs. We surveyed the following distributions for SASL
support: \textit{Gentoo, Debian Wheezy, Ubuntu 14.04.1 LTS, Arch Linux,
Centos~6, OpenBSD~5.6, FreeBSD~10}. We found that only since
2014 \textit{Debian Wheezy} and \textit{Ubuntu 14.04.1 LTS} started to link
against SASL by default. 

There are two options regarding the configuration of the
listen socket, restrictive or global. Restrictive, means a specific IP, e.g., the
localhost IP. \verb+127.0.0.1+, or non restrictive using \verb+0.0.0.0+.
We find that for memcached  only \textit{Debian Wheezy} and \textit{Ubuntu
14.04.1 LTS} use the restrictive default.

We find, not surprisingly that the
documentation is large given the feature set of the
services. Instructions for
enabling the optional security features are hidden in Dynamo, misleading in 
memcached, and clear and simple in redis. It seems that initially the security
documentation of MongoDB was also hidden.   On the 10th of February 2015 more than 40.000
MongoDB databases were unprotected on the Internet. Since then a major update
to the documentation took place~\cite{mongodbsec}.

Indeed, the weaknesses in the security models of NoSQL databases 
are known.
While Srinivas and Nair~\cite{srinivas2015security} conclude
that they are on a good path forward with regard to providing CIA, Okman et
al.~\cite{okman2011security} point out that \emph{``Clearly the future
generations of such DBMSs need considerable development and hardening in order
to provide secure environment for sensitive data which is being stored by
applications (such as social networks) using them.''}.

Binaryedge~\cite{binaryedge2015a} finds that there are still more than 175K unprotected
Redis/MongoDB/Memcache instances in the Internet that can be contacted from any
host.

\smallskip
\noindent{\bf{VNC and the Remote Frame Buffer Protocol:}}
\label{sec:vnc}
Virtual Network Computing (VNC) describes the programs and application providing
the Remote Frame Buffer (RFB) Protocol, RFC6143~\cite{rfc6143} dated 2011. RFB
allows a networked computer to use a graphical user interface on a remote
machine.  It has two common use cases (a) remote support for end-user systems
by support staff and (b) remote administration of physically inaccessible
servers.  The latter is, in particular, used in Linux based
virtualization solutions that provide access to the VGA output of the VMs via
VNC~\cite{hammel2011managing}. The latter is the reason why we consider VNC in
this Section.

To cite from the RFB RFC6143~\cite{rfc6143}: \emph{``The RFB protocol as
defined here provides no security beyond the optional and cryptographically
weak password check described in Section 7.2.2.  In particular, it provides no
protection against observation of or tampering with the data stream.  It has
typically been used on secure physical or virtual networks.''}  However, it can
be enhanced with IPsec or SSH for encryption and some implementations support
plain and certificate based authentication.

Since VNC is used for remote administration, it offers many opportunities
for \mk. An obvious end-user system misconfiguration, weak or no
passwords, opens the end-systems to easy attack. Of course, since these are usually company
provided, they ``should not'' be unprotected in the Internet.
For remote administration of SCADA systems this is worse, as
access to them implies access to industrial control
system~\cite{binaryedge2015b,stevenson2014a}.  

VNC for remote access to virtual machine consoles is often used by 
Linux based virtualization solutions, e.g., libvirt and Xen. 
These enable VNC without authentication by default. Moreover, while VNC is 
typically bound to localhost libvirt offers the ``convenience'' option of using
\verb+0.0.0.0+ by uncommenting a more restrictive binding to localhost. 
If an attacker gets control of a VNC port they can use keyboard commands to reboot
the system~\cite{rfc6143} and boot into single user mode where the credentials
for the administrative account may be changed.

To mitigate this, strong, ideally key-based, authentication and built-in encryption is
needed. Therefore, the protocol must be adjusted. Furthermore, the
implementations should provide these security mechanisms in an easy to use manner.
Shodan lists roughly 600,000 publicly available IPs running VNC. Indeed, roughly
2,000 unprotected VNC instances have been used as the basis of a ``security''
roulette, see the report by Stevenson~\cite{stevenson2014a}.

\smallskip
\noindent{\bf{APIs and Microservices:}}
\label{sec:apis}
Remote procedure calls (RPCs) are a fundamental concept introduced before 1981,
with the goal to make execution of code on a remote machine as simple and
straightforward as on the local
machine~\cite{nelson1981remote,birrell1984implementing}.  Today RPCs are
everywhere, but often hidden behind a different name, examples include JSON-RPC,
XML-RPC, Java-RMI, SOAP, REST. Indeed, one often refers to them as application
programming interfaces (APIs).

APIs are commonly used in, e.g, (a) microservices, (b) configuration interfaces, (c)
mobile application services, and (d) Web applications. 
Microservices~\cite{newman2015building,pratistha2003micro,van2006mobile}
reuse the traditional Unix philosophy of combining ``small, sharp
tools''~\cite{hunt2000pragmatic}.  The communication is delegated to
APIs. An example of a configuration interface is the Docker API which allows
operators to orchestrate applications built in docker containers via Web
services~\cite{docker}. Examples of mobile applications are the various APIs
used in a multitude of Android and iOS
apps~\cite{fiebig2013grindr,polakis2015s,masse2011rest}. Web applications often
rely on client-based javascript code that calls back to APIs on the
server-side~\cite{charland2011mobile,cantelon2014node}.  While APIs often use 
RPCs they do
not have to. Some use HTTP as a transport protocol and are either
RESTfull or use
JSON or XML as basis. Others use plain JSON or plain XMLRPC. Yet another group
defines their own protocol often using binary encoding. 

Some APIs provide some form of authentication, authorization, and
encryption. However, APIs are often used in what the application designer thinks
is a fenced environment. Hence, the typical use case has most security elements
disabled~\cite{van2006mobile,alarcon2010restler}. This is the reason that APIs
are in this class as they basically repeat the misconceptions of \early, namely
to trust every client.


Most attacks against APIs are ones that ``just'' use the service. This can
result in non-intended side-effects due to missing authentication and
authorization, e.g., abusing the docker API, or information leakage, e.g., in
mobile application APIs~\cite{fiebig2013grindr,polakis2015s}, or using the
back-end API rather than the client interface to overcome rate limits against
brute force attacks on the original service~\cite{fallon2015celebgate}.
These attacks are enabled by \mks such as APIs that are Internet-wide accessible
due to (a) holes in the firewall, (b) misconfigured bouncer, and (c) inside
attacks via another compromised microservice~\cite{breen2015java}. 

We again find that, as with almost all new cool ideas, security aspects are
the ones that are considered last. Thus, many APIs come with unclear
documentation and/or global binds for their management APIs, e.g., docker,
tomcat, JBoss~\cite{breen2015java}. Even though the meta-documentation
tells reasonably well that APIs have to be secured, the practice does not
adhere to this goal. Indeed, the problems are known for microservices since
2003~\cite{pratistha2003micro}. Moreover, good practices are known but hardly
followed~\cite{newman2015building}.


\subsection{Discussion---\simplicity}
In this class, we often do not have a single protocol but concepts that are realized by
different protocols that all suffer from the same \mks, e.g., key-value stores
and APIs. Overall, we observe a clear trend towards using HTTP as
application layer protocol, likely because HTTP allows middlebox
traversal~\cite{qazi2013simple}. In theory, this opens up the possibility of
''just securing HTTP'' 
rather than having to secure a large number of Internet
protocols~\cite{richter2015distilling}.  Still, the problems of \mk
remain. Even if an API uses SSL for encryption, it can still be abused
by an attacker~\cite{breen2015java} to disclose information.  The origin
of this trend is the need for complex, yet easy to deploy services.

Apart from delegating security issues to HTTP these modern services again rely
on fencing.  During software and system development, the main focus is on
getting the service ready and not necessarily on security.  Hence,
\emph{``\lbrack{}...\rbrack~Its operation environment is assumed to be
  non-hostile~\lbrack{}...\rbrack''} as stated in the Amazon Dynamo
paper~\cite{decandia2007dynamo}.

The underlying misconception is that services can be securely fenced off as
they are only used within ``internal systems''.  That this can lead to \mks has
already been shown in previous discussions.  Nevertheless, many recent security
incidents have lead to prominent examples of data leakage (key-value stores,
APIs)~\cite{binaryedge2015a}.

The motivation for delegating security to a fenced environment is the notion
that security without fencing is a complex and time consuming task, impossible
to achieve, and making the (backend) services unusable. Hence, the spirit
during development follows the start-up culture, which includes reusing available
protocols as they are (Telnet/VNC).

Fencing techniques have advanced as well, and since 2013 include the
concept of containers~\cite{kim2013practical}. A common \mk is captured by the
following idea: ``If we put a service in its own container in its own dedicated
VM in its own VLAN behind a firewall nothing can happen''. However, since
services have to be reachable they have to allow some access. This access often
turns out to be the entry point for attacks.

Containers, e.g., Docker, are a deployment and testing mechanism.  The
motivation is that traditional means of software distribution do not scale
to the cloud as traditional testing/release procedures do not fit the rapid
development cycle. This often leads to monolithic deployment
of formerly modular software components. This is unmanageable if some component
has to be updated, e.g., to patch a security hole. Furthermore, containers
usually come preconfigured which adds additional \mk traps.

Due to their experience with \complex several developers state that one should
not be too strict about security and that security is ``in the way'' of
innovation. This goes as far as e.g. Ren et al.~\cite{ren2012security} 
claiming that: \emph{``Security and privacy is one fundamental obstacle to cloud
computing's success.''}.

Old mistakes are redone as the default assumption of the software
developer is again ``this is an esoteric scenario and not the intended use
case''.  APIs may release more information than
intended~\cite{fiebig2013grindr,polakis2015s}.  Moreover, the cloud adds the
complexities of multi-party trust and the need for
mutual auditability~\cite{chen2010s}.  Additionally, developer guidelines with
a focus on security are hard to find even if they exist.

Overall, we note that, after having had too complex security and unusable systems,
we are now in a world with easily deployable solutions. However, security
is often outsourced. This results in many \mk traps. 


\section{Summary}
\label{sec:conclusion}
\begin{table*}
\begin{center}
\caption{Summary of all selected representative protocols.}
\label{tbl}
\begin{tabular}{p{1.2cm}r|p{0.45cm}p{0.45cm}p{0.4cm}|lllll|cc|lr|rr}
\toprule
 \multicolumn{2}{c|}{\textbf{Example}} & \multicolumn{3}{c|}{\textbf{\Details}}
& \multicolumn{5}{l|}{\textbf{\Mk traps}}&
\multicolumn{2}{c|}{\textbf{\Docs}} & \multicolumn{2}{c|}{\textbf{\Pub}}& \multicolumn{2}{c}{\textbf{\Usage}} \\ \midrule
                  & \rotatebox{90}{\textbf{Introduced}} &
\rotatebox{90}{\textbf{Authentication}} &\rotatebox{90}{\textbf{Authorization}} &
\rotatebox{90}{\textbf{TLS}}& \rotatebox{90}{\textbf{NoAuth}} &
\rotatebox{90}{\textbf{Credentials}} & \rotatebox{90}{\textbf{Artifacts}} &
\rotatebox{90}{\textbf{Fencing}}& \rotatebox{90}{\textbf{NoUse}}&
\rotatebox{90}{\textbf{Bad Defaults}} & \rotatebox{90}{\textbf{Bad Documentation}} &
\multicolumn{1}{c}{\rotatebox{90}{\textbf{Academia}}} &
\multicolumn{1}{c|}{\rotatebox{90}{\textbf{Lore}}}  &
\multicolumn{1}{c}{\rotatebox{90}{\textbf{Total}}} &
\multicolumn{1}{c}{\rotatebox{90}{\textbf{Affected}}} \\ 

\textbf{\mbox{Early} \mbox{Internet}} & & & & & & & & & & & & &  & \\ \midrule
 \multicolumn{1}{c}{FTP}        &   1971 & \yes & \no  &      & \yes & \yes&      &     &  & \no  & \no  &      & 1999     & 3,000,000$^c$ &  \\
 \multicolumn{1}{c}{TFTP}       &   1981 &      &      &      &      &     &      & \yes&  &      &      & 2015~\cite{macfarlane2015evaluation} & 2006     &  & 600,000$^{a}$~\cite{macfarlane2015evaluation} \\
 \multicolumn{1}{c}{SMTP}       &   1982 & \no  &      & \no  & \yes &     & \yes &     &  &      &      & 2004~\cite{jung2004empirical} & 1995  & 5,600,000$^c$ &  \\
 \multicolumn{1}{c}{DNS}        &   1984 & \na  & \na  &      & \yes &     &
\yes &     &  &      & \no  & 2004~\cite{specht2004distributed} & \at2000     &
10,000,000$^c$ & $>$5,000$^{a}$~\cite{streibelt2013exploring} \\ \midrule

 & & & & & & & & & & & & & &\\ 

\textbf{\mbox{Emerging} \mbox{Threats}} & & & & & & & & & & & & & &\\ \midrule
 \multicolumn{1}{c}{NetFlow}    &\at1990 &     &      &     &    &    & &\yes& &      & \yes &     & 2004 &        &  \\
 \multicolumn{1}{c}{DHCP}       &   1993 & \na &      &     &\yes&    & &\yes& & \yes & \yes &     & 1993 &        &  \\
 \multicolumn{1}{c}{SNMPv2}     &   1993 & \no & \no  &     &    &\yes& &\yes& & \yes & \yes &2014~\cite{rossow2014amplification} & \at2012       &  & 3,800,000$^c$ \\
 \multicolumn{1}{c}{NFSv3}      &   1995 & \no & \no  & \no &\yes&    & &\yes& & \no  & \no  &1999~\cite{harris1999tcp} & 2015 &       100,000$^c$ &  \\
 \multicolumn{1}{c}{Munin}      &\at2000 &     &      &     &    &    & &\yes& &      &      &  &  &        & 6,000$^c$ \\
 \multicolumn{1}{c}{iSCSI}      &   2002 & \no & \no  &     &\yes&    & &\yes& &\yes & \no  &  & 2009 &  & 1,000$^b$\\ \midrule

 & & & & & & & & & & & & & &\\ 

\textbf{\mbox{Complex} \mbox{Security}} & & & & & & & & & & & & & &\\ \midrule
 \multicolumn{1}{c}{IPSec}      &\at1995 & \yes& \yes & \yes&    & &&     &\yes & \no  & \no  & 2000~\cite{ferguson2000cryptographic} & 2005     &        &  \\
 \multicolumn{1}{c}{LDAP}       &   1997 & \no & \no  & \yes&\yes& &&     &     & \no  & \no  &     & 2006     &       200,000$^c$ &  \\
 \multicolumn{1}{c}{IPP}        &   1999 & \no & \no  & \no &\yes& &&     &     & \yes & \yes &     & 2003     &        & 400,000$^c$ \\
 \multicolumn{1}{c}{NFSv4+Krb5} &2000    & \yes& \yes & \yes&    & &&     &\yes &      & \yes &     & 2009     &        &  \\
 \multicolumn{1}{c}{SNMPv3}     &   2002 & \yes& \yes & \yes&    & &\yes&     &\yes & \no  & \no  &     & 2013     &        &  \\
 \multicolumn{1}{c}{IPFix}      &   2008 & \na & \na  & \na &\yes& &&     &     & \no  &  \no &     & 2014     &        &  \\ \midrule

 & & & & & & & & & & & & & &\\ 

\textbf{\mbox{A new} \mbox{Simplicity}} & & & & & & & & & & & & & &\\ \midrule
  \multicolumn{1}{c}{Telnet}     &   1969 & \yes& \no  &     &\yes&\yes& &\yes& & \yes & \yes &2014~\cite{krenc2014internet} & 1993 &       10,600,000$^c$ & 400,000$^{a}$~\cite{carna} \\
  \multicolumn{1}{c}{KV Stores}  &   2006 & \no & \no  & \no &    &    & &\yes& & \no  & \no  & 2011~\cite{okman2011security} & 2015 &           & 175,000$^{a}$~\cite{binaryedge2015a} \\
  \multicolumn{1}{c}{APIs}       &   2000 & \no & \no  & \yes&\yes&    & &\yes& & \yes & \yes & 2003~\cite{pratistha2003micro}    & 2013           & 200,000$^c$ &     \\
  \multicolumn{1}{c}{VNC}        &   1998 & \no &      & \no &\yes&\yes& &\yes& & \no  & \no &2011~\cite{holm2011quantitative} & 2014       & 600,000$^c$ & $>$2,000$^{a}$~\cite{binaryedge2015b} \\ 

\bottomrule
 & & & & & & & & & & & & & &\\ 
 \multicolumn{2}{c|}{} & \multicolumn{3}{p{1.9cm}|}{\mbox{\yes:~Commonly~Used}
\mbox{\no:~Uncommon} \mbox{\na:~Not~Implemented}
\mbox{\hspace{1.5mm}:~Not~Specified}} & \multicolumn{5}{l|}{\yes: \mk traps}&
\multicolumn{2}{p{1.0cm}|}{\mbox{\yes:~Most} \mbox{\no:~Some}} &
\multicolumn{2}{p{1.0cm}|}{\mbox{Year~Published} \mbox{\lbrack example ref.\rbrack}}& \multicolumn{2}{p{0.5cm}}{$^a$~cited~data $^b$~zMap $^c$~Shodan} \\ 

\end{tabular}
\vspace{-0.7cm}
\end{center}

\end{table*}

Our systematization of \mk prone Internet protocols and services
is based on assumptions about the attackers---weak vs.\ strong
attacker---as well as security trade-offs---\good vs.\ \perfect---in the protocols. 
We find that \mk prone Internet protocols and their services fall into one of our four
classes: \early, \emerging, \complex, and \simplicity.

We observe that protocol design and service development have come full circle with regard to
security. In the sense that initially the Internet was a cooperative
environment, see \early. Once it was realized that the Internet was
hostile it was presumed that it was possible to fence of services, see
\emerging. However, attackers are getting stronger and assets more valuable
and, thus, there is the attempt to get security ``right'', see \complex.  The
drawback is complexity and, thus, we complete the circle back to a simple
security model in a 
presumed friendly environment guaranteed by enhanced fencing mechanisms, see
\simplicity. Overall, we find many \mk traps in all of the above classes.

Fencing is seen as a major security solution and often used as only barrier
protecting major assets. However, fencing is a major \mk trap in itself. After
all, with regards to fencing there is the statement from
RFC3365~\cite{rfc3365}: \emph{``History has shown that applications that
  operate using the TCP/IP Protocol Suite wind up being used over the
  Internet. This is true even when the original application was not envisioned
  to be used in a “wide area” Internet environment.''}

We observe that protocol/service designers are clearly forced from one
extreme---full but too complex security---to the other extreme---hardly any
security but easily deployable software that works ``out of the box''.  Indeed,
our observation is that operators get frustrated if the software does not behave
as it should or prevents them from doing what they want to do.  Moreover, for many
decision makers, security is not an end in itself but it is a cost factor that they
``also have to take care of''---if it cannot be avoided. Yet, most
often a security incident proves to be more costly than proactive security 
considerations would have been.

\subsection{Lessons Learned}
\vspace{-0.2cm}
Besides these observations, we compile a set of action points and
requirements protocol designers MUST consider when they
create new or update old protocols.

\noindent
\Early: The major lesson from this class is that security requirements and
environments change. However, this class also demonstrates that such issues can
be tackled. Hence, \textit{\textbf{when updating a protocol, one MUST purge
    problematic use-cases and design choices.}} Only then can appeals for
fixing bad configurations combined with sanctioning of insecure practices lead
to improved overall security (see, e.g.,~\cite{jung2004empirical}
and~\cite{kuhrer2014exit}). While the community succeeded with SMTP, the
Internet still suffers from, e.g., DNS and NTP amplification attacks. In
addition, other protocols from this era still linger and urgently
require revisiting.

\noindent
\Emerging: Given that the non-hostile Internet is gone, a new paradigm
emerged. Instead of adjusting the protocols to the new insecure environment,
the environment is re-defined to be fenced. Thus, the assumption is that the
service is behind a firewall or that lower layers offer security guarantees.
Yet, it is common knowledge that firewalls tend to fail eventually and lower
layers do not hold up to their promises. Thus, \textit{\textbf{when designing a
    protocol, one MUST not assume that it will only be used in the environment
    it was designed for.}} Moreover, \textit{\textbf{when designing a protocol,
    one MUST include authentication, authorization and confidentiality
    preserving methods}}. 

\noindent
\Complex: Here the community tried to address some of the above
recommendations. However, the resulting protocols come with other
complications.  Security features designed for an enterprise setup may not be
appropriate in SoHo scenarios, e.g., recall IPP. IPSec is so complex that it is
not in widespread use. Moreover, implementations do not necessarily
interoperate.  In an attempt to ensure good cryptographic algorithms, SNMPv3
and SSHv1 specified algorithms that were good then but not necessarily
now. Hence this class provides us with the following three lessons.  First,
\textit{\textbf{when designing a protocol, one MUST ensure that the security is
    scalable.}} Scalable in the sense that it is applicable to large as well as
small setups. Even the simplest setup should provide confidentiality,
integrity, and availability by default. Second, \textit{\textbf{when designing
    a protocol, one MUST keep it simple and concise}}, to allow multiple
parties to implement interoperable solutions. Third, \textit{\textbf{when
    designing a protocol, one MUST not mandate the use of specific ciphers but
    one MUST exclude plain and weak algorithms}} since cryptographic algorithms
become obsolete over time.

\noindent
\Simplicity: Overwhelmed by the complexity of the previous protocol generation
a new protocol design trend emerged. Protocols become simple again, and
designer focus on getting things done. New technologies, such as ``Cloud'' and
``SDN'' also bring new paradigms. We find that large enterprises develop
protocols very similar to those of the era of emerging threats while ignoring
lessons learned often for the sake of performance. It works well for the
designed environment, e.g., large enterprises where the perimeter is secure,
but fails in different settings.  The concept of ``everything over HTTP'' leads
to ``protocols'' which ignore and/or misuse basic concepts such as
authentication and in particular authorization. In addition, we find that often
new actors, e.g., the automotive industry, adapt IP based technologies and
re-do old mistakes.  Hence, the most important takeaway from this class is that
\textit{\textbf{when designing a protocol, one MUST take past lessons into
    account}}. While this sounds simple history tells us it is not. Indeed,
almost all \mks pointed out by this survey are covered in even the most basic
security textbooks. Yet, they remain common and repeatedly cause major data
leaks and/or outages. Thus, the prime message of this survey is to finally
apply what we have learned.

\noindent
\textbf{\textit{General Observations:}}
Protocol designers MUST be strict in requiring all
security features from implementations and this must be reflected in the RFCs 
or related standards.
To advance the use of TLS, a decent infrastructure, e.g., 
DANE~\cite{rfc6394,rfc6698} is necessary.
If such an infrastructure is unavailable encryption should at least be performed
opportunistically~\cite{Ylonen1996,RothSR2005}.

Opportunistic encryption raises the question of usability.
Usability has to become a key part of the protocol design process: The goal has to be
to make the protocol secure while designing the system such that
it is simple to comprehend and use. The correct way is either the only
way or at least the easiest way~\cite{balfanz2004search,RothSR2005,bernstein2012security}.

Good examples of such design practices exist in the context of user centric
protocols, e.g., SSH~\cite{Ylonen1996} and the large set of encrypted (mobile)
messaging protocols~\cite{7163029}.  The rise of the latter is motivated by
the---emotional---demands of end-users~\cite{kraus2015analyzing} and
professionals~\cite{mcgregor2015investigating} for easy but secure
communication. Important security features include strong mutual authentication
and opportunistic encryption. The same kind of protocol design is needed for
all Internet protocols and, in particular, those used within the infrastructure.

\balance
\bibliographystyle{IEEEtranS}
\bibliography{paper,bibtex-upload}
\balance

\end{document}